\documentclass[conference]{IEEEtran}
\IEEEoverridecommandlockouts
\usepackage{cite}
\usepackage{amsmath,amssymb,amsfonts}
\usepackage{algorithmic}
\usepackage{graphicx}
\usepackage{textcomp}
\usepackage{xcolor}
\def\BibTeX{{\rm B\kern-.05em{\sc i\kern-.025em b}\kern-.08em
    T\kern-.1667em\lower.7ex\hbox{E}\kern-.125emX}}
\begin{document}

\title{An ADHD Diagnostic Interface Based on EEG Spectrograms and Deep Learning Techniques\\
}

\author{\IEEEauthorblockN{Medha Pappula}
\IEEEauthorblockA{\textit{Children's National Hospital} \\
Washington, DC, USA \\
mpappula@childrensnational.org
}
\and
\IEEEauthorblockN{Syed Muhammed Anwar}
\IEEEauthorblockA{\textit{Children’s National Hospital} \\
Washington, DC, USA \\
sanwar@childrensnational.org}
}

\maketitle

\begin{abstract}
This paper introduces an innovative approach to Attention-deficit/hyperactivity disorder (ADHD) diagnosis by employing deep learning (DL) techniques on electroencephalography (EEG) signals. This method addresses the limitations of current behavior-based diagnostic methods, which often lead to misdiagnosis and gender bias. By utilizing a publicly available EEG dataset and converting the signals into spectrograms, a Resnet-18 convolutional neural network (CNN) architecture was used to extract features for ADHD classification. The model achieved a high precision, recall, and an overall F1 score of 0.9. Feature extraction highlighted significant brain regions (frontopolar, parietal, and occipital lobes) associated with ADHD. These insights guided the creation of a three-part digital diagnostic system, facilitating cost-effective and accessible ADHD screening, especially in school environments. This system enables earlier and more accurate identification of students at risk for ADHD, providing timely support to enhance their developmental outcomes. This study showcases the potential of integrating EEG analysis with DL to enhance ADHD diagnostics, presenting a viable alternative to traditional methods.
\end{abstract}
\vspace{6pt}

\section{Introduction}
Attention-deficit/hyperactivity disorder (ADHD) is a widespread neuro-developmental disorder, estimated to affect about 10\% of the global population [1]. Characterized by challenges in attention control, ADHD interferes with many aspects of daily life functioning and long-term development, manifesting in three predominant subtypes: hyperactive, inattentive, and both combined [2]. Understanding these variations is crucial for an accurate diagnosis and targeted interventions.

\vspace{6pt}
The current diagnostic approach, Diagnostic and Statistical Manual of Mental Disorders, Fifth Edition (DSM-V) relies heavily on behavioral observation, focusing on symptoms present before the age of 12, their interference with various aspects of life, and the exclusion of other explanations for the behavior [3,4]. However, this method has limitations, raising misdiagnosis concerns. This is particularly evident in the late diagnosis of girls, who often exhibit ADHD symptoms in a less hyperactive, more inattentive manner. Boys are more likely to be diagnosed early due to their hyperactivity, highlighting a gender bias in the diagnostic process [5].

\vspace{6pt}
Recent advances in deep learning (DL) methodologies have opened doors to more data-driven forms of assessments [6-8]. Physiological signal analysis offers exciting prospects for more accurate ADHD assessments. One such physiological signal, electroencephalography (EEG), which measures electrical brain activity as neurons fire, has shown promise given the neurodevelopmental nature of ADHD [9]. With the rise of commercial EEG headsets, these models can be used in tandem with digital screening tools. These, especially in school environments, can provide low-cost and accessible screening allowing for students to receive the necessary support earlier [10].

\vspace{6pt}
This research harnesses the power of EEG signals, image processing, and deep learning techniques to develop an advanced ADHD assessment tool. This model is then used to develop a novel screening software targeting the most impacted brain areas identified by the model. This screening software, pairing with commercial EEG headsets, ultimately provides a cost-effective method of accurate screening. The data utilized in this study is taken from a public repository representing EEG data from children having ADHD and control group [11]. The data comprises EEG recordings of variable length from 61 children with ADHD and 60 healthy controls aged 7-12 undergoing visual attention tasks. These signals are then transformed into spectrograms which are fed into a Convolutional Neural Network (CNN), specifically the Resnet-18 architecture [12]. This serves as the base for feature extraction. The major contribution is the use of these features to develop a novel screening system that can be easily implemented in public settings such as schools. 

The rest of the paper is organized as follows. Section II describes the methodology used, Section III shows experimental results, and Section IV concludes the work done, presenting next avenues for research.

\section{Proposed Methodology}

\begin{figure*}[htbp]
\centerline{\includegraphics[scale=0.5]{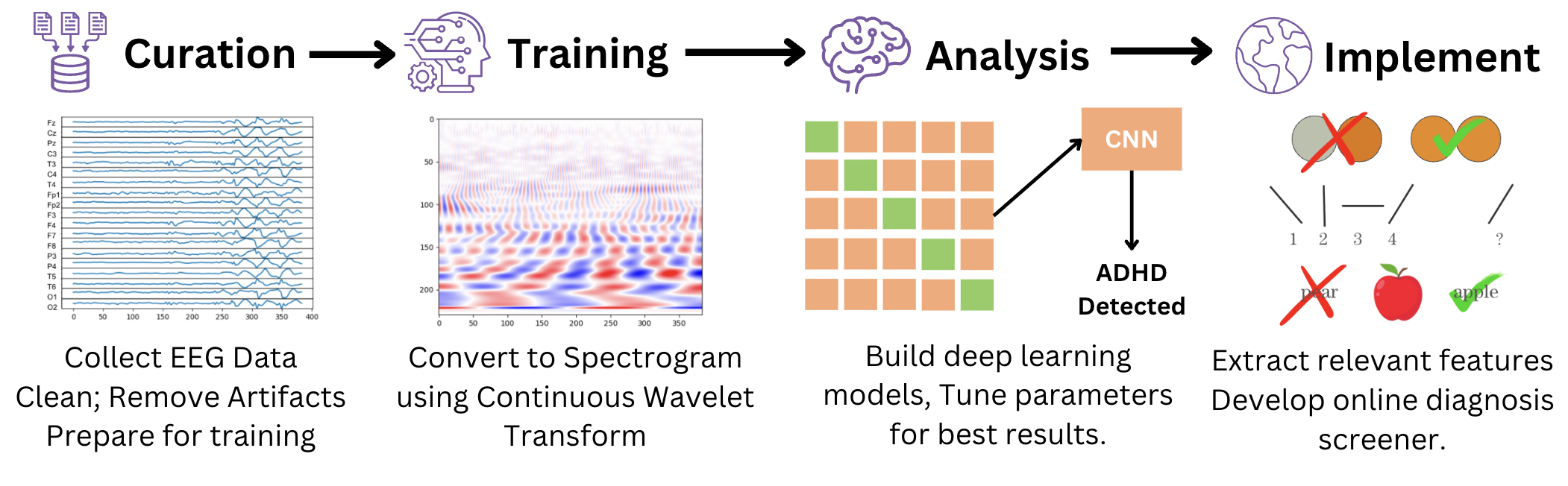}}
\caption{Our proposed approach for ADHD detection using features extracted from EEG spectrograms and deep learning model.}
\label{fig1}
\end{figure*}

The implementation of this project involved four key sections shown in Figure 1. The first step was to collect data from the \textit{EEG data for ADHD / control children} dataset [11]. The core of the methodology includes the training and subsequent feature extraction of the model, utilizing spectrograms to accurately visualize the data and using CNNs, specifically Resnet-18 architecture to develop the classification model. Resnet-18 in particular was used for its ability to accurately interpret the complex patterns hidden within the spectrogram data. The vital step includes taking the results from the feature extraction to develop a digital assessment program. These steps result in a robust system able to accurately assess ADHD in student population. Below is a detailed overview of each section

\subsection{Electroencephalography (EEG) data preparation}

The \textit{EEG data for ADHD / control children} comprises of EEG signals of various lengths from 61 children with ADHD and 60 children without (control), with a total of 121 data points. This EEG data comprised 19 channels (Fz, Cz, Pz, C3, C4, T3, T4, Fp1, Fp2, F3, F4, F7, F8, P3, P4, T5, T6, O1, O2) and was recorded at a frequency of 128 Hz using the 10-20 electrode system. The children in the study were aged 7-12, which is around the recommended age for diagnosis of ADHD in the DSM-V criteria. Those in the control group had no previous history of any other psychiatric disorders, epilepsy, or high-risk behaviors. Each child was subjected to a visual attention task, counting characters on a screen, which is a known area of weakness of those with ADHD [13]. EEG recordings varied in length depending on the response time of each child. All recordings were stored in raw format using a \textit{.mat} file.

\vspace{6pt}
To prepare the data, each of the input \textit{.mat} file was meticulously processed to generate sub-videos. This preparation involved leveraging Python’s MNE and pandas packages. Initially, the EEG data was imported as a pandas DataFrame, with recordings from each individual segmented into distinct subgroups. Following this, an MNE info object was constructed to encapsulate the recording conditions, which were then updated to reflect the accurate experimental settings. The data was subsequently converted into an MNE RawArray, allowing for advanced processing. To enhance signal quality and ensure higher accuracy, a band-pass filter was applied, spanning from 1 to 30 Hz. The filtered data was then segmented into 3-second intervals with a 2-second overlap (1-second overlap on either side), optimizing the temporal resolution for further analysis. This pre-processing ensures that the data is well-structured and ready for subsequent analytical steps.

\subsection{Spectrogram generation}

These overlapping  segments were then separately converted into spectrograms. In particular the Continuous Wavelet Transform (CWT) is a powerful tool for analyzing signals with time-frequency resolution, allowing for a deep examination of how the frequency content of a signal evolves over time. The CWT involves convolving the input signal with a family of wavelet functions, each of which is scaled and translated to capture different frequency components.
\begin{equation}
    \label{eq:cwt}
    CWT_{x}(a, b) = \int_{-\infty}^{\infty} x(t) \cdot \psi^* \left( \frac{t - b}{a} \right) dt
\end{equation}
This process is mathematically expressed in Eq. \ref{eq:cwt} where \textit{x(t)} represents the input signal, \(\psi\)(t) denotes the wavelet function, a is the scale parameter, and b is the translation parameter. In this study, CWT was employed to generate time-frequency representations of EEG signals. The resulting time-frequency maps are of fixed size, which influences the input size of the deep learning model. Fig. 2 shows the spectrograms of binary classified ADHD diagnosis of different subjects. Such an approach for EEG analysis has also been found useful in other tasks such as seizure [14] and motor imagery classification [15]. Utilizing advanced classification algorithms can help in learning identifying specific differences between the condition and control groups.

\begin{figure*}
\centering
\includegraphics[width=1.20\columnwidth]{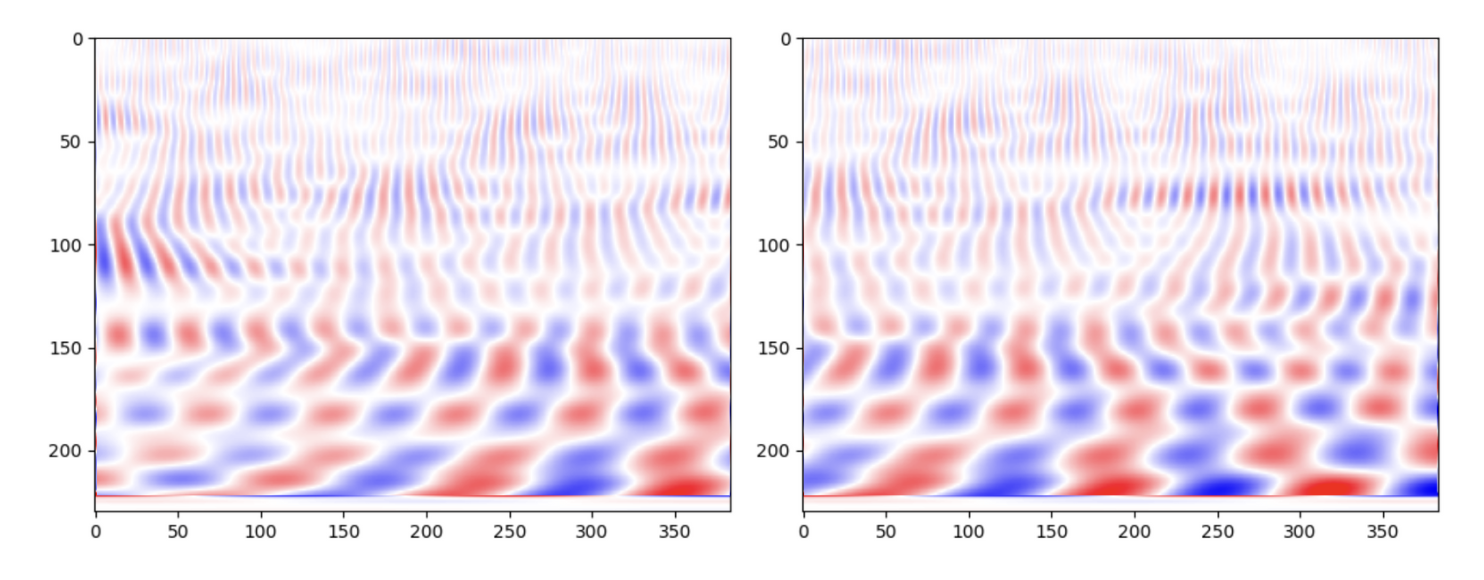}
\caption{Samples of CWT for children with (left) and without (right) ADHD.}
\label{fig2}
\end{figure*}

\subsection{Deep Learning Model}

To extract features from the spectrogram data, a CNN architecture was used. Specifically, this study utilized Resnet-18, which is well known for its ability to capture intricate differences in image data. This model, trained from scratch on the transformed EEG data performed a binary classification on the individual’s ADHD assessment. Each of the total 43153 samples had a structure of 100 time points over 19 channels. A 5-fold cross-validation accuracy was used to evaluate the model's effectiveness. Following this initial training, hyperparameter tuning for both the model and the spectrogram generation further enhanced model ability to perform optimally. The model was used as a baseline as a method of diagnosis. Primarily, the second half of this paper describes its use for feature extraction rather than binary classification. This extraction enables the creation of the front-end interface.

\subsection{Feature Extraction}

To aid in the development of a non-EEG based screening device, the most important features were extracted from the final model. This was done by comparing the baseline and permuted accuracy. Permuted accuracy was determined by replacing a specified feature in the testing data with randomized values and recalculating the model's accuracy. This method demonstrated how significantly the model relied on that feature's data for its predictions. This process was repeated 19 times for each input channel to ensure robust results. As shown in Fig. 3, electrodes FP1/2 (frontal), P3/4 and P7/8 (parietal), and O1/2 (occipital) are found to be significant in ADHD assessment

\begin{figure}[htbp]
\centerline{\includegraphics[scale=0.36]{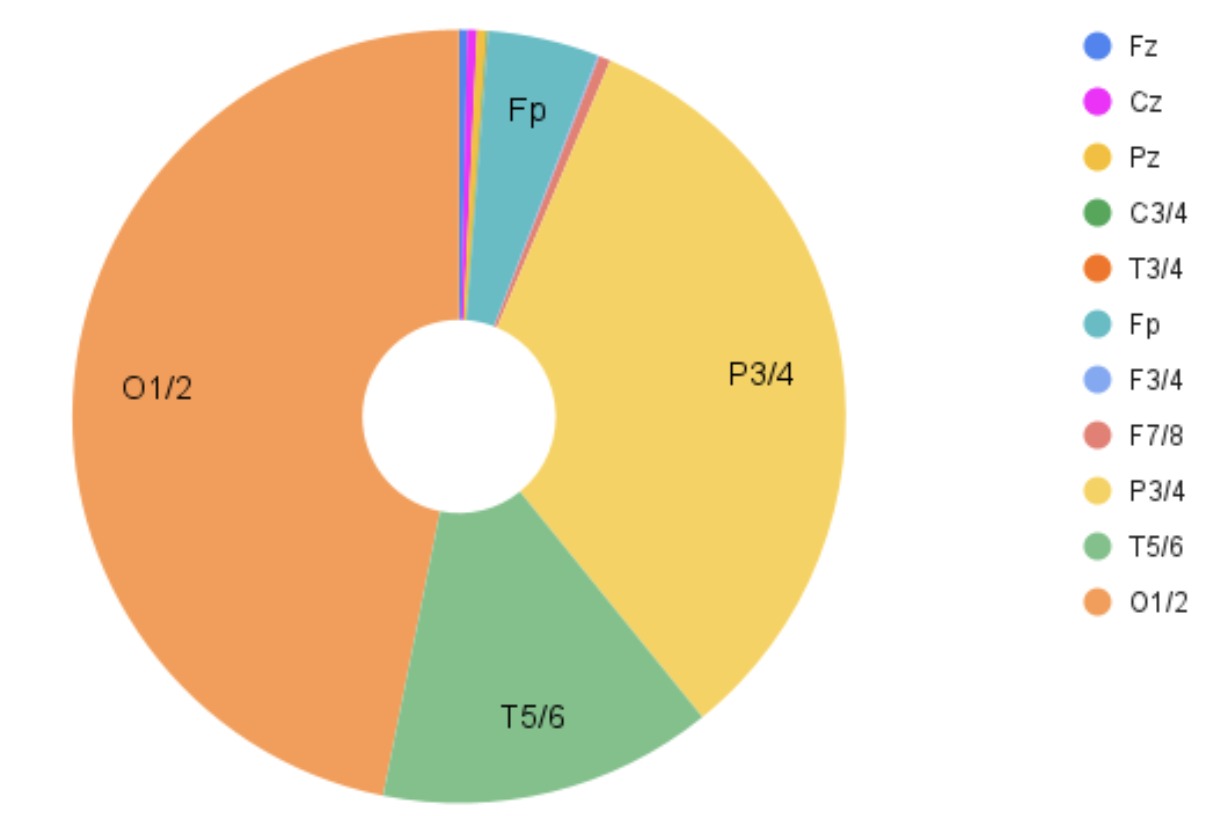}}
\caption{Relative importance of each EEG channel in overall ADHD prediction.}
\label{fig3}
\end{figure}

\section{Experimental Results}

A publicly available \textit{EEG data for ADHD / control children} dataset was used for the binary classification of ADHD prevalence. These recordings were then cleaned and computed into spectrograms. This computation and the training of the deep learning model was performed using the Google's Colaboratory platform. Table 1 presents the precision, recall, and F1-score for the classification with/without ADHD. The results show that Resnet-18 has strong performance given its high precision and recall values, along with a consistent high F1 score. This suggests that Resnet-18 effectively captured the ADHD classification based on EEG data. This 90\% accuracy is similar and at times exceeds other similar studies. For example 86.7\% in looking at objective measures and 68.8\% training a model on medical records.

\begin{table}[htbp]
    \centering
    \caption{Performance metrics for ADHD classification.}
    \begin{center}
    \begin{tabular}{lccc}
        \hline
        & \textbf{Precision} & \textbf{Recall} & \textbf{F1-Score} \\
        \hline
        \textbf{0 (No ADHD)} & 0.98 & 0.79 & 0.88 \\
        \textbf{1 (ADHD)} & 0.86 & 0.99 & 0.92 \\
        \hline
        \textbf{Accuracy} & - & - & 0.90 \\
        \textbf{Macro avg} & 0.92 & 0.89 & 0.90 \\
        \textbf{Weighted avg} & 0.91 & 0.90 & 0.90 \\
        \hline
    \end{tabular}
    \label{table:metrics}
    \end{center}
\end{table}

\begin{figure*}[!t]
\centering
\includegraphics[width=\columnwidth]{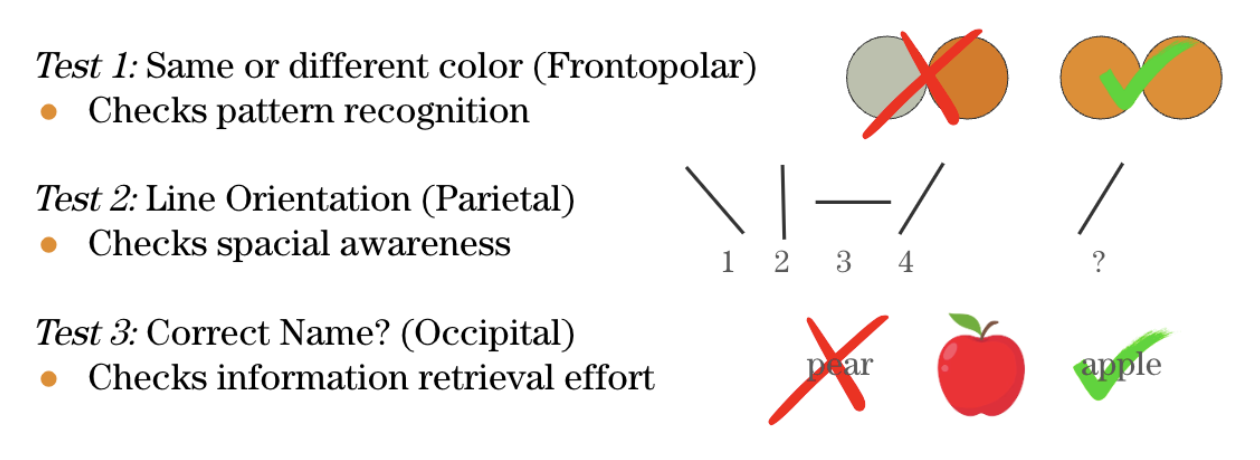}
\caption{Visual representation of the ADHD identification tests.}
\label{fig4}
\end{figure*}

\subsection{Significant EEG channels}

Feature extraction highlights that channels FP1/2 (frontopolar), P3/4 (parietal), P7/8 (parietal), and O1/2 (occipital) are significantly affected in children with ADHD. Changes or damage in these areas are associated with difficulties in focusing on multiple stimuli, making decisions, and perceiving stimuli, which are typical behaviors observed in children with ADHD [16]. This supports current research suggesting decreased gray matter in those regions of the brain [17].

\subsection{Front-end interface}

The results from feature extraction informed the development of a targeted cognitive test system shown in Fig. 4. This system comprises three tests:

\vspace{3pt}
\textbf{Test 1:} Frontopolar Lobe Function- Patients identify whether two circles are the same or different colors.

\vspace{3pt}
\textbf{Test 2:} Parietal Lobe Function- Patients determine the orientation of a line, using a reference orientation map with numbered lines. This assesses spatial awareness.

\vspace{3pt}
\textbf{Test 3:} Occipital Lobe Function- Patients match an image with a word, testing the occipital lobe's ability to retrieve information.

\vspace{3pt}
The reaction time and accuracy can be collected to better inform the degree of the individual's behavior, and can be improved when paired with a commercial EEG headset. This type of interface is easily implementable in school systems allowing for more students to be accurately diagnosed earlier.

\section{Conclusion}

This paper explored the potential use of EEG in ADHD assessment, employing the Resnet-18 architecture on a publicly available dataset. Significant features were extracted by applying CNNs on spectrograms developed from EEG activity. For children with ADHD, the model identified frontopolar, parietal, and occipital areas to be the most impacted. The model had an outstanding performance, with a high F1 score of 0.9. The cognitive testing system based on the identified areas provide a straightforward evaluation, compared to the previous behavior based methods. Overall, this paper demonstrates the feasibility EEG use in ADHD diagnosis and introduces a novel approach to ADHD screening, particularly in settings like schools. In the future we plan to accurately test this front end interface and improve upon it, with the hope to deploy it within public settings.

\end{document}